\documentclass[12pt]{article}
\usepackage{graphicx}
\usepackage{amsmath,amssymb}
\title{  
\vspace*{-2.3cm}  
\begin{flushright}  
\normalsize{  
CERN-PH-TH/2005-131\\
UAB-FT-584}  
\end{flushright}  
\vspace{1.5cm}  
\Large \textbf{\sc Split extended supersymmetry}\vspace*{1.0cm} \author{\large\textbf{Ignatios
Antoniadis~$^{a,}$\footnote{On leave of absence from CPHT, Ecole
Polytechnique, UMR du CNRS 7644}}, \textbf{Karim Benakli~$^b$},
\textbf{Antonio Delgado~$^a$}, \\[0.3cm] \textbf{Mariano
Quir\'os~$^c$} and \textbf{Marc Tuckmantel~$^{a,d}$}\\ \\[0.5cm]
$^a$\normalsize\emph{TH-Division, CERN, 1211 Geneva, Switzerland}\\
$^b$\normalsize\emph{Laboratoire de Physique Th\'eorique et Hautes
Energies}\\ 
\normalsize\emph{Universit\'es de Paris VI et VII, France}\\
$^c$\normalsize\emph{Instituci\'o Catalana de Recerca i Estudis
Avan\c{c}ats (ICREA)}\\ \normalsize\emph{Theoretical Physics
Group, IFAE/UAB, E-08193 Bellaterra, Barcelona, Spain}\\
$^d$\normalsize\emph{Institut f\"ur Theoretische Physik, ETH
H\"ongerberg, 8093 Z\"urich, Switzerland }}}

\date{}
\begin{document}  
\maketitle  
\begin{abstract}

We show how splitting supersymmetry reconciles a class of intersecting
brane models with unification.  The gauge sector in these models
arises in multiplets of extended supersymmetry while matter states are
in $N=1$ representations. A deformation of the angles between the
branes gives large masses to squarks and sleptons, as well as
supersymmetry breaking contributions to other string states. The
latter generate at one-loop heavy Dirac masses for Winos and gluinos
and can induce a mass term for the Higgsino doublets.  We find that
this scenario is compatible with gauge coupling unification at high
scale for both cases where the gauge sector is $N=2$ and $N=4$
supersymmetric.  Moreover a neutralino, combination of neutral
Higgsinos and Binos, is a natural candidate for dark matter.

\end{abstract}
\newpage

\noindent{\sf \underline{INTRODUCTION}} The necessity of a Dark Matter
(DM) candidate and the fact that LEP data favor the unification of the
three Standard Model (SM) gauge couplings are smoking guns for the
presence of new physics at high energies. The latter can take the form
of supersymmetry which, if broken at low energies, offers a framework
for solving the gauge hierarchy problem. Supersymmetry is also welcome
as it naturally arises in string theory, which provides a framework
for incorporating the gravitational interaction in our quantum picture
of the universe.
Recently strong evidence has been accumulated for the presence of a
tiny dark energy in the universe. This raises another hierarchy
problem which is not solved by any known symmetry. It leads to
reconsider our notion of naturalness, possibly also affecting our view
of mass hierarchy. It was then proposed to consider that supersymmetry
might be broken at high energies without solving the gauge hierarchy
problem. More precisely, making squarks and sleptons heavy does not
spoil unification and the existence of a DM candidate while at the
same time it gets rid of all unwanted features of the supersymmetric
SM related to its complicated scalar sector. On the other hand,
experimental hints to the existence of supersymmetry persist since
there are still gauginos and Higgsinos at the electroweak (EW)
scale. This is the so-called split supersymmetry
framework~\cite{split}.  Implementing this idea in string theory is
straightforward~\cite{Antoniadis:2004dt}. However one faces a generic
problem: in simple brane constructions the gauge sector comes in
multiplets of extended supersymmetry~\cite{Blumenhagen:2005mu,marc}.
In this work we show that these economical string-inspired brane
models allow for unification of gauge couplings at scales safe from
proton decay problems. Moreover they provide us with a natural DM
candidate.

\noindent{\sf \underline{BRANE MODELS}} The simplest supersymmetric
brane models are obtained as compactification on a six-dimensional
torus which is a product of three factorized tori. The states in these
models can be assembled into three sets. The first set is made of
strings with both ends on the same stack of branes, leading to $N=4$
vector multiplets. In the following we will also consider a departure
from this minimal case, when two chiral adjoint $N=1$ multiplets are
projected out to remain with $N=2$ vector
supermultiplets\footnote{Note that there are also brane constructions
with no extended supersymmetry in the gauge sector based on
non-toroidal compactifications}. The second set contains strings which
stretch between two stacks of branes that intersect only in two out of
the three internal tori, giving rise to $N=2$ hypermultiplets. The
last set contains strings which are localized at intersection points
of branes in all the three tori and the associated states form $N=1$
supermultiplets.  In this setup of supersymmetric limit, the SM states
are identified as follows:
\begin{itemize}

\item Gauge bosons emerge as massless modes of open strings with both
ends on the same stack of coincident branes. They arise in $N=2$ or
$N=4$ supermultiplets which are decomposed, for each gauge group
factor $G_a$, into one $N=1$ vector superfield $W_a$ and one or three
chiral adjoint superfields $A_a$, respectively.

\item Quarks and leptons are identified with massless modes of open
strings localized at point-like brane intersections and belong to
$N=1$ chiral multiplets.

\item Pairs of Higgs doublets originate from $N=2$ supersymmetry
preserving intersections. They are localized in two tori where branes
intersect, while they propagate freely in the third torus where the
two brane stacks are parallel. Here we assume that all possible
additional non-chiral states that may appear in generic string
constructions can be made superheavy.
 
\end{itemize}

Supersymmetry breaking is then achieved by deforming brane
intersections with a small angle $\Theta$. As a result, a $D$-term
with $\langle D\rangle=\Theta M_S^2$ appears, associated to a
corresponding magnetized $U(1)$ factor with superfield strength
${\mathcal W}$. Here, $M_S$ is the string scale.  Supersymmetry is
then broken and soft masses are induced:
\begin{itemize}

\item A tree-level mass $m_0\propto\sqrt{\Theta}M_S$ for squarks and
sleptons localized at the deformed intersections. All other scalars
acquire in general high masses of order $m_0$ by one loop radiative
corrections. Appropriate fine-tuning is needed in the Higgs sector to
keep light $n_H$ doublets.

\item A Dirac mass~\cite{Fox:2002bu} is induced through the
dimension-five operator
\begin{equation}
\frac{a}{M_S} 
\int d^2\theta {\mathcal W}\, W^a A_a\ \Rightarrow
m^{D}_{1/2}\sim
a \frac{m_0^2}{M_S}\, ,
\label{Dgaugino}
\end{equation}
where $a$ accounts for a possible loop factor.  Actually, this
operator arises quite generally at one-loop level in intersecting
D-brane models with a coupling that depends only on the massless
(topological) sector of the theory~\cite{marc}. Note that this mass
does not break $R$-symmetry and provides an answer to a problem of
split supersymmetry related to the mechanism of generating gaugino
masses.
\end{itemize}


\noindent{\sf \underline{UNIFICATION}} We now study the compatibility
of this framework with one-loop unification. In the energy regime
between the unification scale $M_{GUT}$ and the EW scale $M_W$, the
renormalization group equations meet three thresholds. From $M_{GUT}$
to the common scalar mass $m_0$ all charged states contribute.  Below
$m_0$ squarks and sleptons (which do not affect unification), adjoint
scalars and $2-n_H$ Higgses decouple, while below $m^{D}_{1/2}$ the
$N=2$ or $N=4$ gluinos and Winos drop out.  Finally, at TeV energies
Higgsinos (and maybe the Binos) decouple and we are left at low
energies with the Standard Model with $n_H$ Higgs doublets.  For the
purpose of this computation we use $M_S\sim M_{GUT}$ and we vary $a$
between $a=1$ and $a=1/100$. Realistic values for $M_{GUT}$ and $m_0$
are obtained in both $N=4$ and $N=2$ cases.  The results are
summarized in Table~\ref{table}.

\begin{table}[htb]
\centering
\begin{tabular}{|c|c|c|c|c|c|}
\hline
 &$n_H$ & $a$& $M_{GUT}$  & $m_0$  &$m_{1/2}^D$ \\
 \hline
$N=2$  & $1$& $1$ & $2.8\times 10^{18}$  & $4.5\times 10^{12}$  &
 $7.2\times 10^6$ \\
 & $1$& $1/100$ & $3.8\times 10^{18}$  & $3.2\times 10^{13}$  &
 $2.7\times 10^6$ \\
& $2$ & $1$& $4.5\times 10^{16}$  & $1.1\times 10^{13}$ &
 $2.7\times 10^9$ \\
& $2$ & $1/100$& $4.5\times 10^{16}$  & $8.6\times 10^{13}$ &
 $1.6\times 10^9$ \\
\hline\hline
 $N=4 $& $1$& $1$ & $9.7\times 10^{18}$  & $8.5\times 10^{15}$  &
 $7.4\times 10^{12}$ \\
 & $1$& $1/100$ & $ 10^{19}$  & $6.8\times 10^{16}$  &
 $3.4\times 10^{12}$ \\
 & $2$& $ $ & --- & --- & ---\\
 \hline
\end{tabular}
\caption{\it Values for the unification scale $M_{GUT}$, scalar masses
$m_0$ and Dirac gaugino masses $m_{1/2}^D$ in GeV for $N=2,4$
supersymmetric gauge sector, $n_H=1,2$ light Higgses, and varying the
loop factor $a$.}
\label{table}
\end{table}

Notice that we have imposed perfect unification at one-loop. Although
 this is not necessary from the string theory point of view, it is one of
 the main motivations of split supersymmetry and can be imposed along
 the lines of~\cite{Antoniadis:2004dt}.  In this case, for $N=4$ there
 is \emph{no} solution with $n_H=2$ Higgses at low energies, whereas
 for $N=2$ we find a solution with either $n_H=1$ or $n_H=2$. In all
 cases the unification scale is high enough to avoid problems with
 proton decay. For the two possible cases with one light Higgs ($N=2$
 or $N=4$), $M_{GUT}$ is very close to the Planck scale so that there
 should be no need to explain the usual mismatch between these two
 scales. Varying the loop factor $a$ from 1 to $1/100$ amounts to an
 increase by one order of magnitude in the value of $m_0$, but
 $M_{GUT}$ and $m^{D}_{1/2}$ remain stable within $\mathcal O(1)$
 factors.

The low energy sector of these models contains, besides the SM, just
some fermion doublets (Higgsinos) and eventually two singlets (the
Binos from the discussion below). It therefore illustrates the fact
that only these states are needed for a minimal extension of the SM
consistent with unification and DM candidates, and not the full
fermion spectrum of split supersymmetry.  A similar observation was
recently done in the literature~\cite{Arkani-Hamed:2005yv}, without
the presence of Dirac gauginos associated to the intermediate scale
$m^{D}_{1/2}$.  Here, we do not find these solutions because they do
not unify at one-loop.


\noindent{\sf \underline{DARK MATTER}} Another constraint on the
models is that they must provide a DM candidate.  As usually in
supersymmetric theories this should be the lightest neutralino.  Pure
Higgsinos $(\tilde H_1,\bar{\tilde H}_2)^T$ cannot be DM candidates
because their mass is of Dirac type:~$-\mu \tilde H_1 \tilde H_2+h.c.$
Quasi-Dirac Higgsinos would interact inelastically with matter via
vector-like couplings as $i(\bar{\tilde H}_-\bar\sigma_\mu\tilde H_+ 
-\bar{\tilde
H}_+\bar\sigma_\mu\tilde H_-)$ where $\tilde H_\pm\sim \tilde H_1\pm
\tilde H_2$ are the mass eigenstates with mass eigenvalues
$\mu\pm\epsilon$ respectively. In Dark Matter direct detection
experiments $\tilde H_-$ can only scatter inelastically off of a
nucleus of mass $m_N$ by transitioning to $\tilde H_+$ if $\epsilon
<\epsilon_0
\simeq \frac{1}{2}\beta^2 m_N$~\cite{Smith:2001hy}.  For Ge
experiments with $m_N=73$ GeV and taking a typical escape Dark Matter
particle velocity $\beta c\simeq 600$ km/s one obtains
$\epsilon_0\simeq 146$ keV.  Since direct detection experiments have
ruled out Dirac fermions up to masses of order $50$ TeV some mixing
coming from the Binos is required in order to break the degeneracies
of the two lightest neutralinos and provide a mass difference
$\epsilon >\epsilon_0$ preventing inelastic scattering off the
nucleus.  The mass difference $\epsilon >\epsilon_0$ translates into
an upper bound on the Dirac gaugino mass of about $10^5$ GeV, for the
required Higgsino mass splitting to be generated through the EW
symmetry breaking mixing (of order $m_W^2/m_{1/2}^{D}$) described
below. This value compared to the values in Table~\ref{table} leads to
the $N=2$, $n_H=1$ case as the only possibility to accommodate
it~\footnote{In this case and to destroy the Dirac nature of mass
eigenstates the Higgs sector should be arranged (as quarks and leptons
do) in $N=1$ multiplets.}.  In fact one could have an order of
magnitude suppression of the induced Dirac mass for Binos relative to
the other gauginos, which is not unreasonable to assume in brane
constructions.

In the other two models, the required suppression factor is much
higher and the above mechanism would be very unnatural.  However,
since Binos play no r\^ole for unification as they carry no SM charge,
we could imagine a scenario where $m_{1/2}^{D}$ vanishes identically
for Binos, but not for the other gauginos.  For instance consider the
case where Dirac masses from the operator (\ref{Dgaugino}) are
generated by loop diagrams involving $N=2$ hypermultiplets with
supersymmetric masses of order $M_{GUT}$ and a supersymmetry breaking
splitting of order $\Theta$.  It is then possible to choose these
massive states such that they carry no hypercharge.  In that case
Binos can only have Majorana masses.  Such masses are in general
induced by a dimension-seven effective operator, generated at two-loop
level~\cite{ant}:
\begin{equation}
\frac{b^2}{M_S^3}\int d^2\theta {\mathcal W}^2 {\rm Tr} W^2\Rightarrow
m_{1/2}^M\sim b^2 \frac{m_0^4}{M_S^3}\, ,
\label{majorana}
\end{equation}
where $b$ is another loop factor.  Putting numbers in the above
formula, we get for the $N=4$ $n_H=1$ model $m^M_{1/2}\sim 5\times
10^6$ GeV, which is the upper bound for DM with Majorana Bino
mass~\footnote{This higher value for the upper bound on Majorana
($\sim 5\times 10^{6}$ GeV) versus Dirac Bino mass ($\sim 10^5$ GeV)
we used before is due to the fact that the induced Higgsino mass
splitting through EW symmetry breaking mixing is further suppressed by
the weak angle in the Dirac case.}.  On the other hand, for the $N=2$
$n_H=2$ model, we get $m^M_{1/2}\sim 100$ GeV which is obviously
acceptable. Finally, for the $N=2$ $n_H=1$ case, $m^M_{1/2}\sim 10$
keV which does not play any r\^ole if there is also a Dirac mass, as
we assume and discussed above.  Thus, the constraint of a viable DM
candidate leaves us with two possibilities: (a) $N=2$ with $n_H=1$ and
Dirac masses for all gauginos and (b) $N=4$ with $n_H=1$, or $N=2$
with $n_H=2$ and Majorana mass Binos.


\bigskip

\noindent{\sf \underline{HIGGSINO MASSES}} The Higgsinos themselves
must acquire a mass of order the EW scale.  This is induced by the
following dimension-seven operator, generated at one loop
level~\cite{ant,marc}:
\begin{equation}
\frac{c}{M_S^3}\int d^2\theta \mathcal{W}^2 \overline{D}^2
{\bar H}_1 {\bar H}_2\Rightarrow
\mu \sim c \frac{m_0^4}{M_S^3}\, ,
\label{mu}
\end{equation}
where $c$ is again a loop factor. The resulting numerical value is of
the same order as $m_{1/2}^M$ of Eq.~(\ref{majorana}). Thus, such an
operator can only give a sensible value of $\mu$ for the $N=2$ $n_H=2$
model. In the other two cases, $N=4$ or $N=2$ with $n_H=1$, $\mu$
remains an independent parameter.

Another constraint on these models may come from the life-time of the
extra states. We first consider the case of $N=2$.  Scalars can decay
into gauginos, Dirac gluinos decay through squark loops sufficiently
fast and Dirac Winos and Binos decay into Higgses and Higgsinos.  On
the other hand, for $n_H=2$, there are two Majorana Binos at low
energies.  However, as mentioned above, in the models we consider the
two Higgs doublets form an $N=2$ hypermultiplet. It follows that there
is an $N=2$ coupling between both Binos with Higgses and Higgsinos,
implying that the only stable particle is the usual lightest sparticle
(LSP).  Finally, in the $N=4$ model, scalars still decay into
gauginos, but we have now \emph{two} Dirac gluinos, Winos and Binos;
half of them decay as before, either through scalar loops or into
Higgs-Higgsinos, while the other half can only decay through string
massive states.  Their lifetime is then estimated by
%
$\tau\sim\left(M_S/10^{13}\,\mbox{GeV}\right)^4\,
\left(10^2\, \mbox{GeV}\,/m_{\tilde g}\right)^5\tau_U$,
%
where $m_{\tilde g}$ is the gaugino mass and $\tau_U$ is the lifetime
of the universe.  For gluinos and Winos there is no problem, but Binos
are very long lived although still safe, with a life-time of order
$\tau_U/10$.


To summarize, at low energies we end up with two distinct scenarios
after all massive particles are decoupled: i) $n_H=1$ with light
Higgsinos (models with $N=2$ and $N=4$ gauge sector and $n_H=1$), and;
ii) $n_H=2$ with light Higgsinos and Binos (model with $N=2$ gauge
sector and $n_H=2$). In the $n_H=1$ scenario the DM candidate is
mainly Higgsino, although the much heavier Bino is light enough to
forbid any vector couplings. The relic density can be estimated (using
for example the DarkSUSY program~\cite{Gondolo:2002tz}) and reproduces
the actual WMAP results for $\mu\sim 1.1$ TeV.

The $n_H=2$ scenario is more interesting since there are more
particles at low energies.  The $N=2$ coupling between
Binos-Higgs-Higgsinos leads, after EW symmetry breaking to the
neutralino mass matrix,
\begin{displaymath} 
\left( \begin{array}{cccc} 
M & 0 & m_z s_w c_\beta &  m_z s_w s_\beta  \\ 
0 & M & - m_z s_w s_\beta & m_z s_w c_\beta \\
 m_z s_w c_\beta & - m_z s_w s_\beta& 0 & -\mu \\
 m_z s_w s_\beta & m_z s_w c_\beta & -\mu & 0  \\
\end{array}
\right)
\end{displaymath}
in the basis $({\tilde B}_1, {\tilde B}_2, {\tilde H}_1, {\tilde
H}_2)$ and where $M=m_{1/2}^M$ stands for the Bino Majorana masses.
The mass matrix can be diagonalized to obtain:
\begin{equation}
m_\chi=1/2 \left[(M+ \epsilon_1\mu)-\epsilon_2\sqrt{(M-
\epsilon_1\mu)^2+4 m_z^2 s_w^2} \right]
\end{equation}
where the four different mass eigenvalues are labeled by
$\epsilon_{1,2}=\pm 1$.

The values of $\mu$ and $M$ can then be chosen in such a way as to
reproduce the WMAP observed DM density, as previously
studied~\cite{Giudice:2004tc}. There are then three cases:

\begin{itemize}
\item $M\ll\mu$: the Bino does not interact strongly
enough to annihilate and will in general overclose the universe.

\item $M\gg\mu$: this model converges to the $n_H=1$
scenario and WMAP results require $\mu\sim 1.1$ TeV.
                                                                
\item $M\sim\mu$: the lightest neutralino ($\chi$) is in general a
mixture of Higgsinos and Binos and is a natural candidate for DM. Low
values of $\mu$ are now possible.
\end{itemize}


\noindent{\sf \underline{COLLIDER PHENOMENOLOGY}} In the $n_H=1$
scenario with $\mu\sim 1.1$ TeV, the heavy spectrum will be hardly
observable at LHC while a Linear Collider with center of mass energy
of around 2.5 TeV will be needed to detect a possible signature.

In the $n_H=2$ scenario, the main collider signature is through the
production of charginos.  Their mass is given by $m_{\chi^\pm}= \mu +
\delta \mu$, where $\delta \mu$ is due to electromagnetic
contributions and is of order $300$ to $400$ MeV.  The produced
charginos will decay into the neutralino, mainly through emission of a
virtual $W^{\pm}$ which gives rise to lepton pairs or pions depending
on its energy. This decay is governed by the mass difference $\Delta
m_\chi=m_{\chi^\pm}-m_{\chi^0}$, which is a function of the two
parameters $M$ and $\mu$.  Because charginos are produced through EW
processes, LHC will mainly be able to explore the case of very light
charginos, which exist only in the limited area of the parameter space
with $M\sim \mu$. Unlike in low energy supersymmetry, the absence of
cascade decays in this case will make it difficult to separate the
signal from similar events produced by Standard Model $W^\pm$
production processes.

The search at future $e^+ e^-$ colliders is more promising, and can be
discussed either as a function of the model parameters $(M,\mu)$, or
as a function of the low energy observables $(m_{\chi^\pm}, \Delta
m_\chi)$.  For most of the $(M,\mu)$ parameter range, $\Delta m_\chi$
is small, at most of order a few GeV.  Because the value of $\delta
\mu$ is not small enough to make the chargino long-lived as to produce
visible tracks in the vertex detectors, we have to rely on its decay
products.  This degeneracy implies that the produced leptons or pions
are very soft and it would typically be difficult to disentangle them
from the background due to emission of photons from the beam.  The
strategy is then to look for $e^+ e^- \rightarrow \gamma
+{E_T\hskip-12pt/\hskip4pt}$.  A proper cut on the transverse momentum
of the photon allows to eliminate the background of missing energy due
to emission of $e^+ e^-$ pairs along the beam, as the conservation of
transverse momentum implies now a simultaneous detection of electrons
or positrons~\cite{Chen:1995yu}. The best possible scenario is when
$M$ and $\mu$ are of the same order since, as soon as $M$ starts to be
greater than $\mu$, the Binos quickly decouple and this model
converges to the $n_H=1$ scenario with $\mu\sim 1.1$ TeV. The Higgs sector 
of these models will also give a signal at LHC. The case with just one Higgs at 
low energies predicts a mass $\lesssim 160$  GeV\cite{Giudice:2004tc} depending on
the exact value of $m_0$. In the case of two Higgses, one of them will have this bound
whereas the mass of the other doublet is controlled by $m_A$ as in the MSSM. The result is
that one Higgs will always be discovered at LHC whereas the others will depend
on the value of $m_A$, which is a free parameter. In any case it will be imposible
to distinguish these models from another type of split supersymmetry or 
from a general two Higgs 
model just at LHC. A careful measurement of Higgs coupling will be needed to
disentangle between the different models that could be achieved at the ILC.


\bigskip

\noindent{\sf \underline{CONCLUSION}} Before closing, we would like to
make a few comments concerning the number of parameters and
fine-tuning issues. Dirac gaugino masses, and consequently scalar
masses related through Eq.~(\ref{Dgaugino}), are fixed by the one-loop
unification condition according to Table~\ref{table}. In the two
$n_H=1$ models (with $N=4$ or $N=2$ gauge sector) the Higgsino mass
$\mu$ is fixed by the DM constraint to $\sim 1.1$ TeV. Since the
supersymmetry breaking scale $m_0$ is high, a fine-tuning is needed
like in split supersymmetry to keep one Higgs scalar light. On the
other hand, in the $n_H=2$ model with $N=2$, $\mu$ is determined to
the right scale by the effective operator (\ref{mu}) and the required
Majorana component of Bino mass by (\ref{majorana}).  Obviously in
this case one needs more fine-tuning conditions in order to keep both
eigenvalues of the Higgs mass-squared matrix light. One may wonder
that for a two-by-two symmetric matrix, this implies three conditions
on all its elements. However, it is easy to see that only two
conditions are really needed.  The reason is that the off-diagonal
element is protected by two global low energy symmetries, namely
Pecei-Quinn and $R$-invariance. As a result, if its tree-level value
vanishes, quantum contributions will be proportional to $\mu
m^M_{1/2}\sim{\cal O}({\rm TeV})^2$.


In summary, we presented three viable models of high scale
supersymmetry that naturally emerge in simple string constructions
with intersecting branes and are compatible with gauge coupling
unification and the existence of a dark matter candidate.  Gauginos
come in multiplets of extended supersymmetry and get high scale Dirac
masses by dimension-five effective operators without breaking
$R$-symmetry, consistently with gauge coupling unification. The low
energy sector contains besides the Standard Model particle content a
Higgsino pair providing, in general through mixing with the Bino, a
natural Dark Matter candidate.

\bigskip

This work was supported in part by the European Commission under the
RTN contract MRTN-CT-2004-503369, in part by CICYT, Spain, under
contracts FPA2001-1806 and FPA-2002-00748, in part by the INTAS
contract 03-51-6346 and in part by IN2P3-CICYT under contract Pth
03-1.

\end{document}